\documentclass[12pt]{elsart}
\usepackage{feynmp}
\usepackage{epsfig}
\usepackage{amsmath}
\usepackage{amssymb}
\usepackage{amsfonts}
\newcommand{\cO}{\mathcal{O}} 
\newcommand{\Dd}[1]{\overset{\leftrightarrow}{D}_{#1}}
\newcommand{\Pn}{\psi}
\newcommand{\Pb}{\bar{\psi}}
\newcommand{\dbar}{\bar{d}}
\newcommand{\ubar}{\bar{u}}
\newcommand{\sbar}{\bar{s}}
\newcommand{\rmT}{{\rm T}}
\newcommand{\cal}{\mathcal}
\newcommand{\gev}{{\, \rm GeV}}
\newcommand{\ci}{{\mathrm i}}

\begin{document}
\begin{frontmatter}
\title{
\vspace{-1.9cm}
\hfill {\normalsize \mdseries DESY 01-036} \\[-0.2cm] 
\hfill {\normalsize \mdseries HU-EP-01/10} \\[0.4cm] 
A lattice evaluation of four-quark operators in the nucleon}
\author[Reg]{M.~G\"ockeler},
\author[NIC,HU]{R.~Horsley},
\author[NIC,FU]{B.~Klaus},
\author[NIC]{D.~Pleiter},
\author[Reg]{P.E.L.~Rakow},
\author[Reg]{S.~Schaefer},
\author[Reg]{A.~Sch\"afer} and
\author[NIC,DESY]{G.~Schierholz}

\address[Reg]{Institut f\"ur Theoretische Physik,
    Universit\"at Regensburg, D-93040 Regensburg, Germany} 
\address[NIC]{John von Neumann-Institut f\"ur Computing NIC, D-15735 Zeuthen,
Germany}
\address[HU]{Institut f\"ur Physik, Humboldt-Universit\"at zu Berlin,
                     D-10115 Berlin, Germany}  
\address[FU]{Institut f\"ur Theoretische Physik,
  Freie Universit\"at Berlin, D-14195 Berlin, Germany}
\address[DESY]{Deutsches Elektronen-Synchrotron DESY, D-22603 Hamburg, Germany}
 
\date{}
  
\begin{abstract}
Nucleon matrix elements of various four-quark operators are evaluated in
quenched lattice QCD using Wilson fermions. Some of these operators 
give rise to twist-four contributions to nucleon structure functions.
Furthermore, they bear valuable information about the diquark structure
of the nucleon. Mixing with lower-dimensional operators is avoided
by considering appropriate representations of the flavour group.
We find that for a certain flavour combination of baryon structure
functions, twist-four contributions are very small. This suggests that
twist-four effects for the nucleon might be much smaller than
$m_p^2/Q^2$.
\end{abstract}

\end{frontmatter}

\section{Introduction}

The knowledge of higher quark and gluon correlators in hadrons is of
fundamental interest in order to understand the structure of baryons 
and mesons on the basis of QCD. Matrix elements of four-quark operators
contain information on the quark and diquark structure of the nucleon.
Within the operator product expansion (OPE),
four-quark operators give rise to higher-twist contributions 
(cat's-ear diagram). While this has been known for many years 
\cite{webber,ellis,bag,wilco,wilco2,mira}, 
the size of these contributions is still uncertain. Because
the structure function  $F_2(x,Q^2)$ of the proton is one of the best 
measured hadronic quantities, the natural idea would be to extract 
the higher-twist contribution from the $Q^2$ dependence of $F_2$. This has,
however, proven to be a difficult task (for a recent attempt 
see Ref.~\cite{Alekhin:2000ch}). 

A computation of higher-twist effects from first principles is possible 
by means of Monte Carlo simulations of lattice QCD, and first estimates 
using this method in the case of the pion became available recently 
\cite{pion}. In this paper we shall extend our previous work to the 
case of the nucleon.

By means of the OPE  $F_2(x,Q^2)$ can be expressed
through forward nucleon matrix elements of local operators. 
In the deep inelastic limit
$Q^2 \to \infty$ it is dominated by the leading twist-two
contributions. These have been the subject of intensive studies in the
past. The next-to-leading contributions have twist four and are
suppressed by one power of $1/Q^2$. 
More precisely, the OPE relates (Nachtmann) moments \cite{nachtmann}
$ \int_0^1 \mathrm dx \, x^{n-2} F_2(x,Q^2) \big|_{\mathrm
  {Nachtmann}} $, which take into account the effects of the finite
proton mass $m_p$,  to
the product of  Wilson coefficients  and hadronic
matrix elements. Schematically one finds for $n=2,4,6,\ldots$ 
\begin{eqnarray} 
    & \int_0^1 & \mathrm dx \, x^{n-2} F_2(x,Q^2) \big|_{\mathrm
      {Nachtmann}} \nonumber \\ 
     & \equiv &
    \int_0^1 \frac{\mathrm dx}{x^3} \xi ^{n+1} F_2(x,Q^2)\big [n^2+2n+3
    +3(n+1)(1+4 m_p^2 x^2 / Q^2)^{1/2} 
    \nonumber \\  & & {} + n (n+2) 4 m_p^2 x^2 / Q^2
    \big ] / (n+2)(n+3)   \nonumber \\ 
    & = & c^{(2)}_n(Q^2/\mu^2,g(\mu)) A^{(2)}_n (\mu)
    + \frac{c^{(4)}_n(Q^2/\mu^2,g(\mu))}{Q^2} A^{(4)}_n (\mu)
    + O \left(\frac{1}{Q^4}\right) 
\end{eqnarray}
with $\xi = 2x/(1+\sqrt{1+4 m_p^2 x^2 / Q^2})$.

The reduced matrix elements $A^{(t)}_n$ of twist $t$ and spin $n$ depend
on the renormalisation scale $\mu$. The mass dimension of 
$A^{(t)}_n$ is $t-2$. The dimensionless Wilson coefficients  $c^{(t)}_n$ can be
calculated in perturbation theory. In the flavour-nonsinglet
channel, the twist-two operators are two-quark operators,
\begin{equation}
  \Pb\gamma_{\mu_1} \Dd{\mu_2} \cdots \Dd{\mu_n} \Pn \,,
\end{equation}
symmetrised in all indices and with trace terms subtracted.

The four-quark operators we are interested in have twist four and higher.
In particular, the twist-four, spin-two matrix element $A^{(4)}_2$ is given
by  (indices in $\{ \dots \}$ are symmetrised)
\begin{equation} 
  \frac{1}{2} \sum_S \langle P, S | A^c_{\{\mu \nu \}} 
  - \mbox{trace} |   P, S \rangle  \equiv
  \langle P | A^c_{\{\mu \nu \}} 
  - \mbox{trace} | P \rangle
  = 2 A^{(4)}_2 (P_\mu P_\nu -  \mbox{trace} ) \label{opt3}
\end{equation}
with the four-quark operator
\begin{equation} \label{opt4}
  A^c_{\mu \nu } = (\Pb G \gamma_\mu \gamma_5 t^a \Pn) 
  (\Pb G \gamma_\nu \gamma_5 t^a \Pn)  
\end{equation}
(using the nomenclature introduced in Ref.~\cite{pion}).
The quark field $\Pn$ carries flavour, colour, and Dirac indices, the
matrices $t^a$ are the usual generators of colour SU(3)$_{\mathrm c}$, 
and for two flavours the flavour matrix $G$ reads 
\begin{equation} 
  G = \mbox{diag} ( e_u , e_d ) = \mbox{diag} ( 2/3 , -1/3 ) 
\end{equation}
in terms of the quark charges $e_q$.
The proton states with momentum $P$ and spin vector $S$ are normalised 
such that 
\begin{equation}
  \langle P, S | P', S' \rangle = (2 \pi)^3 2 E_P \delta({\mathbf P} -
  {\mathbf P'}) \delta_{S S'} \label{normal}
\end{equation}
The Wilson coefficient reads~\cite{wilco,wilco2} 
$c_2^{(4)} = g^2 \left( 1+O(g^2) \right)$.

These expressions are to be compared with their twist-two counterparts:
\begin{equation} 
  \langle P  | \cO _{\{\mu \nu \}} 
  - \mbox{trace} | P \rangle 
  = 2 A^{(2)}_2 (P_\mu P_\nu -  \mbox{trace} ) 
\end{equation}
with the operator
\begin{equation} \label{opt2}
  \cO _{\mu \nu} =   
  \frac{\ci}{2} \Pb G^2 \gamma_\mu \Dd{\nu} \Pn 
\end{equation}
and the Wilson coefficient $c_2^{(2)} = 1+O(g^2)$.

The operators (\ref{opt4}) and (\ref{opt2}) transform identically under
Lorentz transformations, but (\ref{opt4}) has dimension six, whereas 
(\ref{opt2}) has only dimension four: four-quark operators will in general
mix with two-quark operators of lower dimension. This fact complicates
the investigation of four-quark operators, because the mixing with 
lower-dimensional operators cannot be calculated reliably within
perturbation theory. 
A nonperturbative computation in lattice QCD could proceed along the
same lines as in the case of the twist-three matrix element $d_2$
\cite{Gockeler:2000ja}. 
 For the time being, we do not attempt such
a nonperturbative calculation of the renormalisation and mixing
coefficients of four-quark operators. Instead we restrict ourselves
to cases where mixing with lower-dimensional operators is prohibited
by flavour symmetry.

We shall present results obtained in the quenched approximation of
lattice QCD with Wilson fermions. A preliminary account of some of
these results has already been given in Ref.~\cite{Capitani:2000je}. Since
the lattice formulation of gauge theories requires an analytic
continuation from Minkowski space to Euclidean space, we now switch to 
the Euclidean formulation. In particular, all operators will be written 
down in Euclidean space, unless otherwise noted.

\section{The general framework}

In our previous publication \cite{pion} we have studied four-quark 
operators in the
pion. In this case we could avoid mixing with lower-dimensional
operators by working with operators which carry isospin
$I=2$. Obviously, operators with $I=2$ vanish in the proton. 
Therefore we enlarge the flavour symmetry group from SU(2)$_{\mathrm F}$
to SU(3)$_{\mathrm F}$
assuming three quarks of the same mass.  
Correspondingly, we then have 
\begin{equation} 
  G = \mbox{diag} ( e_u , e_d, e_s ) = \mbox{diag} ( 2/3 , -1/3, -1/3 ) 
\end{equation}
and the flavour structure of the operator in the OPE is now
\begin{equation}
  \cO = (e_u \ubar u + e_d \dbar d + e_s \sbar s)
  (e_u \ubar u + e_d \dbar d + e_s \sbar s) \,. \label{OPE_op}
\end{equation}
While two-quark operators transform under SU(3)$_{\mathrm F}$ according
to $\overline{\mathbf{3}} \otimes \mathbf{3} = \mathbf{1} \oplus \mathbf{8}$,
we have for four-quark operators: $(\overline{\mathbf{3}} \otimes \mathbf{3})
\otimes (\overline{\mathbf{3}} \otimes \mathbf{3}) 
= 2 \cdot \mathbf{1} \oplus 4 \cdot \mathbf{8} \oplus
\mathbf{10} \oplus \overline{\mathbf{10}} \oplus \mathbf{27}$.
Four-quark operators with $I=0,1$, $I_3 = 0$, and hypercharge $Y=0$ 
belonging to the multiplets
$\mathbf{10}$, $\overline{\mathbf{10}}$, $\mathbf{27}$
do not mix with two-quark operators and do not automatically 
vanish in a proton expectation value. The operators belonging to the 
$\mathbf{27}$ multiplet are (giving only the flavour structure) 
\begin{equation}
  \label{op27}
  \begin{split}
    \cO^{I=1}_{\bf{27}} =& 
    \frac{1}{10} (e_u^2 -e_d^2 -2 e_u e_s +2 e_d e_s)  [ 
    (\ubar u) (\ubar u) - (\dbar d) (\dbar d) 
    \\
    &- (\ubar s) (\sbar u) - (\sbar u) (\ubar s) 
    + (\dbar s) (\sbar d) + (\sbar d) (\dbar s) 
    \\
    &- (\sbar s) (\ubar u) - (\ubar u) (\sbar s) 
    + (\sbar s) (\dbar d) + (\dbar d) (\sbar s) ] \ ,
  \end{split}
\end{equation}

\begin{equation}
  \label{op27I0}
  \begin{split}
    \cO^{I=0}_{\bf{27}} =&  \frac{1}{60}(e_u^2+e_d^2+e_u e_d -3 e_u
    e_s -3 e_d e_s +3 e_s^2) 
    \\&
    [ 2( \ubar u) (\ubar u) +2 (\dbar d) (\dbar d) 
    +(\dbar d)(\ubar u)+(\dbar u)(\ubar d)
    \\&
    +(\ubar d)(\dbar u)+(\ubar u)(\dbar d) 
    - 3(\ubar s) (\sbar u) - 3(\sbar u) (\ubar s)
    \\&
    -3 (\dbar s) (\sbar d) -3 (\sbar d) (\dbar s) 
    -3 (\sbar s) (\ubar u) - 3(\ubar u) (\sbar s)
    \\&
    -3 (\sbar s) (\dbar d) -3 (\dbar d) (\sbar s) 
    +6 (\sbar s)(\sbar s) ] \ .
  \end{split}
\end{equation}
Inserting the values of the quark charges one finds
\begin{equation}
  e_u^2 -e_d^2 -2 e_u e_s +2 e_d e_s=e_u^2+e_d^2+e_u e_d -3 e_u
  e_s -3 e_d e_s +3 e_s^2 =1 \ .
\end{equation}
As the operators belong to the same multiplet, the Wigner--Eckart
theorem tells us that the proton matrix elements
of these two operators are proportional to each other:
\begin{equation}
  \langle P | \cO^{I=0}_\mathbf{27} |  P \rangle = 
  \frac{1}{2} \langle P  | \cO^{I=1}_\mathbf{27} | P  \rangle \,.
  \label{eq.two27}
\end{equation}
Furthermore, the Wigner--Eckart theorem relates proton matrix elements  
to neutron matrix elements. Thus our results can easily be rephrased in terms 
of neutron expectation values. However, unless otherwise stated, we shall
only present the proton results.

The operators belonging to the multiplets ${\bf 10}$ and 
$\overline{\bf{10}}$ read
\begin{equation}
  \begin{split}
    \cO_{\mathbf{10}}^{I=1} =&(\dbar d)(\ubar u)-(\dbar u)(\ubar d)  
    +(\ubar d )(\dbar u)-(\ubar u)(\dbar d)  \\&
    + (\ubar u) (\sbar s)+( \dbar  s)(\sbar d)-(\dbar d)(\sbar s)
    -(\ubar s)(\sbar u) \\
    &- (\sbar d)(\dbar s) +(\sbar u)(\ubar s) + (\sbar s)(\dbar d)
    -(\sbar s)(\ubar u) \ ,
  \end{split}
  \label{op10}
\end{equation}
\begin{equation}
  \begin{split}
    \cO_{\overline{\mathbf{10}}}^{I=1} =
    &(\dbar d)(\ubar u)+(\dbar u)(\ubar d)
    -(\ubar d )(\dbar u)-(\ubar u)(\dbar d)
    \\ &+ (\ubar u) (\sbar s) -( \dbar   s)(\sbar d)
    -(\dbar d)(\sbar s) +(\ubar s)(\sbar u) \\
    &+ (\sbar d)(\dbar s) 
    -(\sbar u)(\ubar s) + (\sbar s)(\dbar d)
    -(\sbar s)(\ubar u) \ .
  \end{split}
  \label{op10bar}
\end{equation}
Being antisymmetric with respect to the interchange of the two
quark-anti\-quark pairs they do not appear in the flavour decomposition
of the OPE operator (\ref{OPE_op}).

In the following we present Monte Carlo data from quenched simulations
at $\beta = 6.0$ with Wilson fermions on a $16^3 \times 32$ lattice. 
We have performed simulations at three different values of the hopping
parameter $\kappa =0.1515$, $0.1530$ and $0.1550$ and we have collected
about 300 configurations. The statistical errors have been determined by
the jackknife method.

\begin{figure}
\begin{center}
    \epsfig{file = 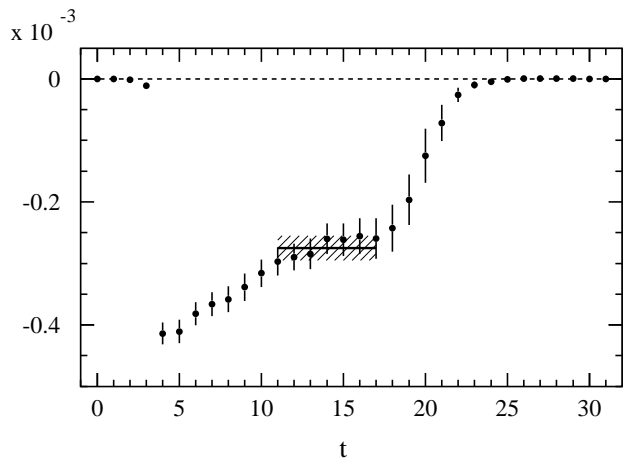, width= 9 cm} 
    \caption{Plateau for the bare matrix element of
	$\cO_{\mathbf{27}}^{I=1}$ for the operator 
        $V^c_{44} - \mbox{trace}$ (see Eq.~(\ref{eq:defop})),
        divided by $m_p^4$, with $\kappa = 0.1515$. \vspace{1cm} }
  \label{fig.plat}
\end{center}
\end{figure}

The proton matrix elements are computed in the standard fashion from 
ratios of three-point functions, $\langle B(t) \cO (\tau) \bar{B}(0) \rangle$,
to two-point functions, $\langle B(t) \bar{B}(0) \rangle$, 
with the interpolating fields $B$ and $\bar{B}$ of Ref.~\cite{Gockeler:1996wg}.
For $ 0 \ll \tau \ll t $ this ratio should be independent of $t$: 
\begin{equation}
  R = \frac{\langle B(t) \cO(\tau) B(0) \rangle}{\langle B(t) B(0) \rangle} = 
  \frac{1}{(2 \kappa)^2 2 m_p}\langle P | \cO | P \rangle  + \cdots 
\end{equation}
If we vary $t$ keeping $\tau$ fixed we should therefore find a 
region where $R$ is constant, i.e.\ shows a plateau. 
An example of such a plateau is shown in Fig.~\ref{fig.plat}.
We have always chosen $\tau = 5$ and
the spatial components of the proton momentum have been set to zero.
The ratio $R$ equals the matrix element in the lattice normalisation
of states and fields. In order to obtain the fields in the continuum
normalisation we have to multiply each quark field by 
$\sqrt{2 \kappa}$. To normalise the states according to Eq.~(\ref{normal})
we must multiply $R$ by an additional factor of $2 m_p$.
We determine the matrix elements by fitting
the ratio $R$ to a constant in the region  $11 \le t \le 17$.

For a general four-quark operator the three-point function 
$\langle B(t) \cO (\tau) \bar{B}(0) \rangle$ consists of three types
of contributions, which can be represented pictorially by the following
diagrams 
\begin{center}
  \begin{fmffile}{fd1}
    \begin{fmfgraph}(60,40) 
      \fmftop{i1} \fmfbottom{o1}
      \fmf{phantom}{i1,v} \fmf{plain}{v,v}  \fmf{plain,left=90}{v,v} 
      \fmf{phantom}{v,o1}
      \fmfdot{v} 
      \fmfforce{(0.5w,0.8h)}{v}
      \fmfforce{(0.1w,0.3h)}{v1}
      \fmfforce{(0.9w,0.3h)}{v2}
      \fmfv{decoration.shape=circle,decoration.filled=shaded,
        decoration.size=0.1h}{v1}
      \fmfv{decoration.shape=circle,decoration.filled=shaded,
        decoration.size=0.1h}{v2}
      \fmf{plain,right=.4}{v2,v1} \fmf{plain,left=.4}{v2,v1}
      \fmf{plain}{v2,v1} 
    \end{fmfgraph} \end{fmffile} 
  \begin{fmffile}{fd2}
    \begin{fmfgraph}(60,40) 
      \fmfleft{i1}  \fmfright{o1} \fmf{phantom}{i1,v} \fmf{phantom}{v,o1}
      \fmfforce{(0.5w,0.54h)}{v}
      \fmfforce{(0.1w,0.3h)}{v1}
      \fmfforce{(0.9w,0.3h)}{v2}
      \fmfv{decoration.shape=circle,decoration.filled=shaded,
        decoration.size=0.1h}{v1}
      \fmfv{decoration.shape=circle,decoration.filled=shaded,
        decoration.size=0.1h}{v2}
      \fmf{plain,left=.4}{v2,v1} \fmf{plain}{v2,v1} 
      \fmf{plain,right=.2}{v2,v,v1}
      \fmfdot{v} 
      \fmf{plain,tension=1.4}{v,v}
    \end{fmfgraph} \end{fmffile} 
  \\
  \begin{fmffile}{fd3}
    \begin{fmfgraph}(60,40) 
      \fmfforce{(0.5w,0.54h)}{v}
      \fmfforce{(0.1w,0.3h)}{v1}
      \fmfforce{(0.9w,0.3h)}{v2}
      \fmfv{decoration.shape=circle,decoration.filled=shaded,
        decoration.size=0.1h}{v1}
      \fmfv{decoration.shape=circle,decoration.filled=shaded,
        decoration.size=0.1h}{v2}
      \fmf{plain,left=.4}{v2,v1} 
      \fmf{plain,left=.4}{v2,v}  \fmf{plain,right=.4}{v2,v} 
      \fmf{plain,left=.4}{v,v1}  \fmf{plain,right=.4}{v,v1} 
      \fmfdot{v} 
    \end{fmfgraph} \end{fmffile} 
\end{center}
It is precisely through contributions of the form of the first two diagrams
that the mixing with lower-dimensional operators occurs. Therefore these
contributions cancel in the operators which we consider, and we are left
with the contributions of the last type only.
For proton matrix elements only some terms of the operators
contribute, e.g. the $(\dbar d)(\dbar d)$ terms  vanish as those
containing $s$ quarks. Therefore the expectation values of the operators
(\ref{op27}), (\ref{op10}), (\ref{op10bar}) reduce to
\begin{equation}
  \begin{split}
    \langle P |\cO^{I=1}_{\bf{27}} | P \rangle&= 
    \frac{1}{10}
    \langle P |
    (\ubar u) (\ubar u)  | P \rangle  \big|^{\bf 27}\,, \\
    \langle P, S |\cO^{I=1}_{\bf{10}} | P, S \rangle&=
    \langle P, S |(\dbar d)(\ubar u)-(\dbar u)(\ubar d)  
    +(\ubar d )(\dbar u) -(\ubar u)(\dbar d)  
            | P, S \rangle \big|^{\bf 10}\,,\\
    \langle P, S |\cO^{I=1}_{\overline{\bf{10}}} | P, S \rangle&=
    \langle P, S |(\dbar d)(\ubar u)+(\dbar u)(\ubar d)  
    -(\ubar d )(\dbar u) -(\ubar u)(\dbar d)  | P, S \rangle 
    \big|^{\overline{\bf 10}}\,.
  \end{split}
  \label{redten}
\end{equation}

\begin{figure}
  \begin{center}
    \epsfig{file=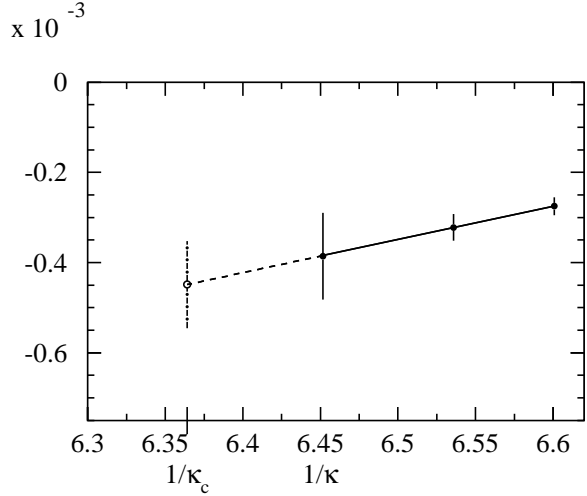,width=9cm} 
    \caption{Chiral extrapolation of the bare proton matrix element of 
      $\cO^{I=1}_{\mathbf{27}}$ for the operator $V^c_{44} - \mbox{trace}$
      (see Eq.~(\ref{eq:defop})), divided by $m_p^4$.\vspace{1cm}}  
      \label{fig.chiex}
  \end{center}
\end{figure}

At each $\kappa$ value the matrix elements are made dimensionless 
by dividing them by the corresponding value of $m_p^4$.
As the bare quark mass is given by $a m_q = 1/(2 \kappa) - 1/(2
\kappa_{c})$
 we perform the chiral limit by linear extrapolation in $1/\kappa$ to 
$1/\kappa=1/\kappa_{c} =6.3642 $.
As an example of our results we show in Fig.~\ref{fig.chiex} the 
chiral extrapolation of the 
bare proton matrix element of $V^c_{44} - \mbox{trace}$ 
(${I=1}$ component in the $\mathbf{27}$ representation of 
SU(3)$_{\mathrm F}$) divided by $m_p^4$. (For the definition of 
$V^c_{\mu \nu}$ see Eq.~(\ref{eq:defop}).)

\section{Operators from the $\mathbf{27}$ multiplet 
	\label{twentyseven}}

The twist-four contribution in the $F_2$ structure function comes from
the four-quark operator $A_{\mu \nu}^c$, see Eq.~(\ref{opt4}). In order to
access the flavour-${\mathbf{27}}$ component experimentally one has to 
combine  the structure functions of several 
baryons  ($p$, $n$, $\Lambda$, $\Sigma$, $\Xi$) in such a way as to
project out the desired flavour combination, e.g.
\begin{eqnarray}
  \langle p | \cO^{I=1}_\mathbf{27} | p \rangle &=& 
  \langle \Sigma^+ | \cO | \Sigma^+ \rangle 
  -2 \langle \Sigma^0 | \cO | \Sigma^0 \rangle 
  + \langle \Sigma^- | \cO | \Sigma^- \rangle  \nonumber
  \\ 
  &=& 
  -\langle \Sigma^+ | \cO | \Sigma^+ \rangle 
  - \langle \Sigma^- | \cO | \Sigma^- \rangle \nonumber
  \\ &&
  -6 \langle \Lambda  | \cO | \Lambda \rangle 
  +2 \langle \Xi^0 | \cO | \Xi^0 \rangle 
  +2 \langle \Xi^- | \cO | \Xi^- \rangle \nonumber
  \\ &&
  +2 \langle p | \cO | p \rangle 
  +2 \langle n | \cO | n \rangle   \,.
\end{eqnarray}
Unfortunately most of these terms will not be measured in the foreseeable 
future. So a direct comparison with data is out of question. On the
other hand, they can be used as a testing ground for models of hadrons,
taking the role of experimental data. Note that the ${\mathbf{27}}$
contribution can also be isolated by studying combinations of 
electromagnetic and weak structure functions \cite{gottlieb}.

Of course, we need to know the renormalised operators. Although due to our
choice of the flavour-${\mathbf{27}}$ component mixing with two-quark
operators is absent, different four-quark operators may still mix under
renormalisation. Therefore we have computed the matrix elements of the 
following operators (using the 
nomenclature introduced in Ref.~\cite{pion}):
\begin{equation}
  \begin{split}
    V^c_{\mu\nu}  &= \Pb G\gamma_{\mu} t^a \Pn 
    \, \Pb G\gamma_{\nu} t^a \Pn ,\\[0.5ex]
    A^c_{\mu\nu}  &= \Pb G\gamma_{\mu} 
    \gamma_5 t^a \Pn \,  \Pb G\gamma_{\nu} \gamma_5 t^a
    \Pn, \\[0.5ex] 
    T^c_{\mu\nu} &= \Pb G\sigma_{\mu\rho} 
    t^a \Pn \,  \Pb G\sigma_{\nu\rho} t^a \Pn,\\[0.5ex] 
    V_{\mu\nu} &= \Pb G\gamma_{\mu} \Pn 
    \,  \Pb G\gamma_{\nu} \Pn ,\\[0.5ex]
    A_{\mu\nu} &= \Pb G\gamma_{\mu} 
    \gamma_5 \Pn \,  \Pb G\gamma_{\nu} \gamma_5 \Pn ,\\[0.5ex]
    T_{\mu\nu} &= \Pb G\sigma_{\mu\rho} 
    \Pn \,  \Pb G\sigma_{\nu\rho} \Pn.\\
    \label{eq:defop} 
  \end{split}
\end{equation}

The bare expectation values divided by $m_p^4$ and extrapolated 
to the chiral limit are given in Table~\ref{tab:op27s2} for the 
the spin-two components, while the traces are given in 
Table~\ref{t6}. E.g.\ the number shown for the operator 
$A^c_{\mu\nu} - \mbox{trace}$ in Table~\ref{tab:op27s2} is what we
obtain for $\langle P |\frac{1}{10} (\ubar \gamma_4 \gamma_5 t^a u)
 (\ubar \gamma_4 \gamma_5 t^a u) -\mbox{trace} | P \rangle / m_p^4$
in the chiral limit.
We have checked that these operators fulfil their Fierz identities.

\begin{table} 
    \caption{Matrix elements of the spin-two operators from the 
    ${\mathbf{27}}$ multiplet, divided by $m_p^4$ and
    extrapolated to the chiral limit.} \label{tab:op27s2}
\vspace{0.5cm} \begin{center}
  \begin{tabular}[h]{ll} \hline
     \multicolumn{1}{c}{operator} & 
    \\ \hline 
    $ A^c_{\{\mu \nu\}}-\mbox{trace}$ & $ (-0.9 \pm 0.8)\cdot10^{-4}$ \\
    $ V^c_{\{\mu \nu\}}-\mbox{trace}$ & $ (-4.5 \pm   1.0)\cdot10^{-4}$\\
    $ T^c_{\{\mu \nu\}}-\mbox{trace}$ & $ (4.9 \pm 0.8)\cdot10^{-4} $ \\
    $ A_{\{\mu \nu\}}-\mbox{trace}$ & $(-0.0 \pm 1.3)\cdot10^{-4} $\\
    $ V_{\{\mu \nu\}}-\mbox{trace}$ & $ (8.4 \pm 1.4)\cdot10^{-4} $ \\
    $ T_{\{\mu \nu\}}-\mbox{trace}$ &  $(-4.6 \pm   1.9)\cdot10^{-4}$ \\ 
  \hline
  \end{tabular}  
\vspace{1.0cm} \end{center}
\end{table}

\begin{table} 
    \caption{Matrix elements of the spin-zero operators from the
    ${\mathbf{27}}$ multiplet, divided by $m_p^4$ and
    extrapolated to the chiral limit.} \label{t6} 
\vspace{0.5cm} \begin{center}
  \begin{tabular}[h]{lll} \hline
     Dirac structure  & \multicolumn{1}{c}{$t_a \otimes t_a$} & 
    \multicolumn{1}{c}{$1 \otimes 1$}\\ \hline 
    $ \gamma_5 \otimes \gamma_5 $ & 
    $(2.8\pm  0.7)\cdot 10^{-4} $ & $(-13.4 \pm  2.6)\cdot 10^{-4} $   \\
    $1 \otimes 1  $ & 
    $(-7.1 \pm  1.2)\cdot 10^{-4} $  & $  (17.8 \pm  2.9)\cdot 10^{-4}$\\
    $\gamma_\mu \gamma_5 \otimes  \gamma_\mu \gamma_5   $ &
    $(17.2 \pm    2.0)\cdot 10^{-4} $ & $ (-12.1 \pm 3.9)\cdot 10^{-4} $\\
    $\gamma_\mu \otimes \gamma_\mu $
    & $ (-18.2 \pm 2.2)\cdot 10^{-4} $& $(7.4 \pm 4.0)\cdot 10^{-4} $ \\
    $\sigma_{\mu \nu} \otimes \sigma_{\mu \nu} $ & 
    $ (-4  \pm  7)\cdot 10^{-4} $ & $(17.6 \pm 9.6)\cdot 10^{-4} $  \\
   \hline
  \end{tabular}
\vspace{1.0cm} \end{center} 
\end{table}

The renormalisation constants have been calculated in one-loop
perturbation theory \cite{pion}.
The renormalised spin-two piece of the operator $A^c_{\mu \nu}$ reads
\begin{equation}
  \begin{split}
    \big[ A^c_{\mu\nu}(\mu) \big]^{\mathrm {ren}}
     = A^c_{\mu\nu} - \frac{g_0^2}{16 \pi^2} \big[ 
    &\left( 3 \ln (a \mu) + 46.072285 \right) A^c_{\mu\nu} \\ 
    & + \left( - \frac{8}{9} \ln (a \mu) + 0.083982 \right) V_{\mu\nu} \\
    & + \left( - \frac{5}{3} \ln (a \mu) + 0.157467 \right) V^c_{\mu\nu} \\ 
    & - 1.071448 \,\, T_{\mu\nu}
    - 2.008965 \,\, T^c_{\mu\nu} \big] \,,
  \end{split}
\end{equation}
where $g_0$ is the bare coupling constant ($\beta \equiv 6/g_0^2$).The
renormalisation scale $\mu$ will be identified with the inverse
lattice spacing $1/a$. 
In our simulations this has a value of $1/a \approx 2.12 \gev$ (using
$r_0$ to set the scale). 
In terms of the renormalised operator the
reduced matrix element $A_2^{(4)}$ is then given by
\begin{equation} 
  \frac{1}{m_p^2}A^{(4)}_2\big{|}^{\rm {\bf 27}, I=1} = \frac{2}{3} 
  \frac{\langle P |\frac{1}{10}
  (\ubar \gamma_4 \gamma_5 t^a u)(\ubar \gamma_4 \gamma_5 t^a u) -\mbox{trace}
    | P \rangle}{m_p^4} \,,
\end{equation}
and we obtain for the lowest moment of $F_2$ in our special flavour channel
\begin{equation} 
    \int_0^1 \mathrm dx \, F_2(x,Q^2) \big|_{\mathrm {Nachtmann}}
    ^{\mathbf{27}, I=1}
    = - 0.0005(5) \frac{m_p^2 \alpha_s (Q^2)}{Q^2} + O(\alpha_s^2) \,.
  \label{nuclres}
\end{equation}
The analogous result for the neutron differs from the above only by
the sign.
                        
In the proton the corresponding twist-two contribution is about
$0.14$ at $Q^2 = 5 \gev^2$. 
As in the pion, the twist-four correction is tiny.
Our result  may be compared with bag model estimates. In this model
the scale for the prefactor in Eq.~(\ref{nuclres}) is set by
$ B/m_p^4 \approx 0.0006$, 
where $B \approx (145 \mbox{MeV})^4$ is the bag constant. The factor
$B/m_p^4$ is however multiplied by a relatively large (and negative) 
number \cite{bag}.
    
It is rather difficult to determine the first moment of the
higher-twist contribution to $F_2(x)$
experimentally. Phenomenological fits to the
available data  give a positive value of about $0.005(4) \gev^2/Q^2$
\cite{Alekhin:2000ch,Choi:1993cu}. 
Our matrix element, which is due to its flavour
structure only one contribution to the full moment, 
is considerably smaller than this phenomenological number.

\section{Operators from the $\mathbf{10}$ and $\mathbf{\overline{10}}$ 
  multiplets}
Having found rather small matrix elements for our four-quark operators
from the $\mathbf{27}$ one may ask if operators from the $\mathbf{10}$
or  ${\overline{\mathbf{10}}}$ of $\mathrm{SU}(3)_{\mathrm F}$
(although not contributing to $F_2$ in the OPE) would have  larger
 matrix elements. With the
two possible colour structures that can form colour singlet operators,
these operators are linear combinations of terms of the form
 $(\Pb G\, \Gamma t^a \Pn)(  \, \Pb G'\, \Gamma' t^a \Pn)$ and 
 $(\Pb G\, \Gamma  \Pn)(  \, \Pb G'\, \Gamma'  \Pn)$, respectively,
where $\Gamma$ and $\Gamma'$ are Dirac matrices.
We have chosen the flavour matrices $G$, $G'$ such that we get the 
following flavour structures:
\begin{equation}  
  (\dbar d) (\ubar u) - (\ubar u) (\dbar d) 
  \label{nostern}
\end{equation}
and 
\begin{equation} 
  (\dbar u) (\ubar d) - (\ubar d) (\dbar u) \ .
  \label{stern} 
\end{equation}
These can be  combined to yield 
the  $\mathbf{10}$ and $\overline{\mathbf{10}}$ structures in
Eq.~(\ref{redten}).

Discrete symmetries impose restrictions on the matrix elements of
these operators.
We have
\begin{equation}
  \langle P, S | \cO | P,S \rangle^* = 
  \langle P, -S | {\cal{T P}}\cO{\cal{P}}^{-1} {\cal T}^{-1} 
        | P,-S \rangle
  = \langle P,S | \cO^\dagger | P,S \rangle
  \label{eq.symm}
\end{equation}
where ${\cal P}$ is the parity and ${\cal T}$ is the time inversion operator.
For the Dirac matrices used in our computations we define sign factors
$s_1$, $s_1'$, $s_2$ and $s_2'$ by
\begin{equation}
  \begin{array}{ccc}
    \gamma_4 \Gamma^\dagger \gamma_4 = s_1 \Gamma &,&
    \gamma_4 {\Gamma'}^\dagger \gamma_4 = s_1' \Gamma' \ , \\
    \gamma_4 \gamma_5 C \Gamma^* C^{-1} \gamma_5 \gamma_4 = s_2 \Gamma &,&
    \gamma_4 \gamma_5 C {\Gamma'}^* C^{-1} \gamma_5 \gamma_4 = s'_2
    \Gamma' \,. 
  \end{array}
\end{equation}
Here $C$ is the charge conjugation matrix with 
$C \gamma_\mu ^\rmT C^{-1} = -\gamma_\mu$.
One more sign $\epsilon_\cO$ is determined by
\begin{equation}
  \langle P, -S | \cO | P,-S \rangle = 
  \epsilon_\cO \langle P, S | \cO | P,S \rangle \ .
\end{equation}
From Eq.~(\ref{eq.symm}) we now get for the flavour structure 
(\ref{nostern})
\begin{equation}
  \langle P, S | \cO | P,S \rangle^* = 
  \epsilon_\cO s_2 s_2'  \langle P, S | \cO | P,S \rangle = 
  s_1 s_1' \langle P, S | \cO | P,S \rangle
\end{equation}
and for the flavour structure (\ref{stern})
\begin{equation}
  \langle P, S | \cO | P,S \rangle^* = 
  \epsilon_\cO s_2 s_2'  \langle P, S | \cO | P,S \rangle = 
  -s_1 s_1' \langle P, S | \cO | P,S \rangle \ .
\end{equation}
Thus the matrix element $\langle P, S | \cO | P,S \rangle$ 
is real if $\epsilon_\cO s_2 s_2' = 1$ and purely imaginary if  
$\epsilon_\cO s_2 s_2' = -1$; the matrix element vanishes
if $\epsilon_\cO s_2 s_2' = -s_1 s_1'$
for the flavour structure (\ref{nostern}) or 
$\epsilon_\cO s_2 s_2' = s_1 s_1'$ for the 
flavour structure (\ref{stern}). We have checked that these
restrictions are satisfied by our results within statistical
errors. We restrict ourselves in the following to the matrix elements
which are not forced to be zero by the above relations. Note that for
a given Dirac structure at most one of the flavour structures
(\ref{nostern}) and (\ref{stern}) yields a non-vanishing result.

The definite Lorentz transformation properties of our operators could be
used to define reduced matrix elements, e.g. in Minkowski space one gets
\begin{equation}
  \langle P,S | (\dbar \gamma_\mu \gamma_5 d)(\ubar \gamma_\nu \gamma_5 u)-
  (\ubar \gamma_\mu \gamma_5 u)(\dbar \gamma_\nu \gamma_5 d) | P,S \rangle =
  A \epsilon_{\mu \nu \alpha \beta} ( P^\alpha S^\beta -S^\alpha
  P^\beta) \,.
\end{equation}
Thus in this case the matrix element with $\mu=1$, $\nu=2$ and $S^\alpha =
\delta_{\alpha 3}$ is equal to 
the one with $\mu=2$, $\nu=3$ and $S^\alpha =
\delta_{\alpha 1}$. 
This holds only on average, so in order to increase the statistics
we averaged over these matrix elements to reduce the statistical error. 
The bare expectation values divided by $m_p^4$ are given together with
their statistical errors in Tables \ref{t1} and \ref{t3}.

\begin{table}  
   \caption{Expectation values of operators with the flavour structures
    (\ref{nostern}) and (\ref{stern}) in an unpolarised proton,
    divided by $m_p^4$ and extrapolated to the chiral limit.} \label{t1}
\vspace{0.5cm} \begin{center}
  \begin{tabular}[h]{lcll} \hline
    Dirac structure & flavour &\multicolumn{1}{c}{$t_a \otimes t_a$}& 
   \multicolumn{1}{c}{$1 \otimes 1$}
    \\ \hline
    $ 1 \otimes \gamma_4$ & (\ref{nostern}) &
    $ (0.4  \pm 0.6)\cdot10^{-3}$& $(3.0\pm  1.6)\cdot10^{-3}$ \\
    $ \epsilon_{4 \alpha \beta \delta}   \gamma_\alpha \gamma_5 \otimes
    \sigma_{\beta \delta}$ &(\ref{nostern})  
    &$(-5.6 \pm  2.8)\cdot10^{-3} \ci$ & $(5.9 \pm  3.6)\cdot10^{-3}\ci$  \\
    $ \gamma_5 \otimes \gamma_4 \gamma_5$ &(\ref{stern})  
    &  $(0.6 \pm  0.5)\cdot10^{-3} $ &  $ (-3.3 \pm 1.8)\cdot10^{-3}$\\
    $\gamma_\alpha \otimes  \sigma_{4 \alpha}$ & (\ref{stern})& 
    $(-3.1 \pm 1.2)\cdot10^{-3}\ci$ & $(0.1 \pm  1.9)\cdot10^{-3}\ci$\\
  \hline
  \end{tabular}
\vspace{1.0cm} \end{center} 
\end{table}

\begin{table} 
    \caption{Expectation values of operators with the flavour structures
    (\ref{nostern}) and (\ref{stern}) in a polarised proton 
    (${\mathbf S} = {\mathbf e}_3$), divided by $m_p^4$ and 
    extrapolated to the chiral limit.} \label{t3}
\vspace{0.5cm} \begin{center}
  \begin{tabular}[h]{lcll} \hline
    Dirac structure & flavour & \multicolumn{1}{c}{$t_a \otimes t_a$} &
    \multicolumn{1}{c}{$1 \otimes 1$}
    \\ \hline 
    $1 \otimes \gamma_3 \gamma_5$  &(\ref{nostern})  &
    $(-3.3 \pm  0.6)\cdot10^{-3} \ci$ & $(15.9 \pm  1.6)\cdot10^{-3} \ci$ \\
    $1 \otimes \sigma_{2 1} $ &(\ref{nostern})  &
    $(-4.1 \pm  0.6)\cdot10^{-3}$ & $(10.7 \pm 1.8)\cdot10^{-3}$\\
    $ \gamma_5 \otimes \sigma_{4 3}$ &(\ref{nostern})  & 
    $(2.2 \pm  0.5)\cdot 10^{-3} $& $ (-4.4 \pm 1.3)\cdot10^{-3}$ \\
    $ \gamma_4 \otimes \gamma_3 \gamma_5- \gamma_3 
    \otimes \gamma_4 \gamma_5$ &(\ref{nostern})  &
    $(-9.1 \pm 0.9)\cdot10^{-3} \ci$ & $(13.0 \pm 2.1)\cdot10^{-3}\ci$ \\ 
    $\epsilon_{3 \alpha \lambda \rho}   \gamma_\alpha  \otimes
    \sigma_{\lambda \rho}$ &(\ref{nostern})  & 
    $(42 \pm 4)\cdot10^{-3}$ & $(-51 \pm  5)\cdot10^{-3}$ \\
    $\gamma_5 \otimes \gamma_3$ &(\ref{stern})  &
    $(4.9 \pm 0.5)\cdot10^{-3}\ci$& $(-15.6 \pm 1.4)\cdot10^{-3}\ci$\\
    $\gamma_2 \otimes \gamma_1$ &(\ref{stern})  &  
    $(2.4 \pm 0.5)\cdot10^{-3} \ci$&$(-5.6 \pm  0.9)\cdot10^{-3}\ci$ \\
    $ \gamma_2 \gamma_5 \otimes \gamma_1 \gamma_5$ &(\ref{stern})  &
    $(6.1 \pm 0.8)\cdot10^{-3} \ci$ & $(-7.6 \pm 1.1)\cdot10^{-3} \ci$ \\
    $\gamma_\alpha \gamma_5 \otimes  \sigma_{3 \alpha}$ &(\ref{stern})  & 
    $ (21.5 \pm  1.8)\cdot10^{-3}$ & $(-19.2 \pm  2.6)\cdot10^{-3}$  \\
    $\sigma_{2 \alpha} \otimes  \sigma_{1 \alpha}$ &(\ref{stern})  &
    $(-8.9 \pm 1.5)\cdot10^{-3} \ci$ & $(8.1 \pm 1.8)\cdot10^{-3} \ci$\\
  \hline                              %
  \end{tabular}
\vspace{1.0cm} \end{center} 
\end{table}

The order of magnitude of the results does not differ greatly from
those found for the operators in the $\mathbf{27}$. The
renormalisation constants for the $\mathbf{10}$ and
$\overline{\mathbf{10}}$ operators are not known, but we do not expect 
that the renormalised operators have much larger matrix elements than
the bare ones.

\section{Diquarks}
The four-quark operators can be rewritten to look like a
diquark density.
We have computed matrix elements of operators of the following form:
\begin{equation}
  \frac{1}{10}(\ubar_a \Gamma \gamma_5   C  \ubar_b^\rmT)
 ( u_{a'}^\rmT C^{-1} \gamma_5 \Gamma'  u_{b'} )
  (\delta_{a b'} \delta_{b a'}-\delta_{a a'} \delta_{b b'}) \,,
  \label{dreiquer}
\end{equation} 
\begin{equation}
   \frac{1}{10}(\ubar_a \Gamma  \gamma_5  C  \ubar_b^\rmT)
 ( u_{a'}^\rmT C^{-1} \gamma_5 \Gamma'   u_{b'} ) 
 (\delta_{a b'} \delta_{b a'}+\delta_{a a'} \delta_{b b'}) \,,
  \label{sechs}
\end{equation}
where $a$, $b$, $a'$ and $b'$
are the colour indices. These are the two possibilities to form a
colour singlet. In Eq.~(\ref{dreiquer}) the diquark is in a
$\overline{\mathbf 3}$ of colour and thus anti-symmetric in colour. 
In Eq.~(\ref{sechs}) it is in a ${\mathbf 6}$ and symmetric in colour.
Because of the Pauli principle it has to be symmetric (anti-symmetric) 
in the other indices. 
The flavour structure being symmetric, the Dirac structure has 
therefore to be symmetric (anti-symmetric). Thus a given Dirac
structure will appear only for one of the two possible colour structures.
The expectation values of operators of the form (\ref{dreiquer}) and
(\ref{sechs}) can be  computed from those of the four-quark operators 
studied in Section~\ref{twentyseven}. 
But in order to get the correct errors we have redone the analysis. The
results (again divided by $m_p^4$) are presented in Tables~\ref{d1}
and \ref{d2} 
for the spin-zero and the spin-two contributions, respectively.

\begin{table} 
    \caption{Matrix elements of the spin-zero operators from the 
    $\mathbf{27}$ multiplet in the diquark picture, divided by $m_p^4$ and
    extrapolated to the chiral limit.} \label{d1}
\vspace{0.5cm} \begin{center}
  \begin{tabular}[h]{lcl} \hline
    \multicolumn{1}{c}{operator}  & colour  \\ \hline 
    $\frac{1}{10}(\ubar \gamma_\mu \gamma_5   \gamma_5 C \ubar^\rmT)
    ( u^\rmT C^{-1} \gamma_5 \gamma_\mu \gamma_5 u) $ 
    & $\overline{\mathbf 3}$ & $(-83 \pm  8)\cdot 10^{-4}$\\ 
    $\frac{1}{10}(\ubar \sigma_{\mu \nu} \gamma_5 C  \ubar^\rmT)
    ( u^\rmT C^{-1} \gamma_5 \sigma_{\mu \nu} u)$
    & $\overline{\mathbf 3}$ 
    & $(42 \pm  22)\cdot 10^{-4} $ \\
    $\frac{1}{10}(\ubar \gamma_\mu \gamma_5  C  \ubar^\rmT)
    ( u^\rmT C^{-1} \gamma_5 \gamma_\mu u)$&${\mathbf 6}$ & 
    $(45 \pm  9)\cdot 10^{-4} $  \\
    $\frac{1}{10}(\ubar \gamma_5 \gamma_5  C  \ubar^\rmT)
    ( u^\rmT C^{-1} \gamma_5 \gamma_5 u)$&${\mathbf 6}$ & 
    $(-0.8 \pm  2.6)\cdot 10^{-4} $  \\
    $\frac{1}{10}(\ubar \gamma_5  C  \ubar^\rmT)
    ( u^\rmT C^{-1}  \gamma_5 u)$&${\mathbf 6}$ & 
    $(-4.9 \pm  6.7)\cdot 10^{-4} $ \\
   \hline
  \end{tabular}
\vspace{1.0cm} \end{center}
\end{table}

\begin{table} 
    \caption{Matrix elements of the spin-two operators from the
    $\mathbf{27}$ multiplet in the diquark picture, divided by $m_p^4$ and
    extrapolated to the chiral limit.} \label{d2} 
\vspace{0.5cm} \begin{center}
  \begin{tabular}[h]{lcl} \hline
    \multicolumn{1}{c}{operator}  & colour  \\ \hline 
    $\frac{1}{10}( \ubar  \gamma_4 \gamma_5 \gamma_5 C  \ubar^\rmT)
    ( u^\rmT  C^{-1} \gamma_5 \gamma_4 \gamma_5 u)-\mbox{trace}$&
    $  \overline{\mathbf 3}$ &
    $ (12.9  \pm  2.2)\cdot 10^{-4}$ \\
    $\frac{1}{10} (\ubar  \sigma_{4 \alpha} \gamma_5 C  \ubar^\rmT)
    (u^\rmT C^{-1} \gamma_5   \sigma_{4 \alpha}  u)-\mbox{trace}$ &
    $  \overline{\mathbf 3}$ &
    $ (-16.8 \pm 4.3)\cdot 10^{-4}$ \\
    $\frac{1}{10}(\ubar \gamma_4 \gamma_5 C  \ubar^\rmT) 
    (\ubar^\rmT C^{-1} \gamma_5  \gamma_4 u)-\mbox{trace}$ &
    ${\mathbf 6}$ &
    $(3.9 \pm  2.9)\cdot 10^{-4}$ \\
  \hline
  \end{tabular}
\vspace{1.0cm} \end{center} 
\end{table}

Strictly speaking, we are again studying operators from the
${\mathbf{27}}$ representation of ${\mathrm{SU}}(3)_{\mathrm F}$,
whose $\ubar \ubar u u$ component is given by Eqs.~(\ref{dreiquer}) and
(\ref{sechs}), respectively.  At least within the quenched
approximation it seems however reasonable to consider (\ref{dreiquer}) and
(\ref{sechs}) as representing valence diquark densities. In the same
spirit, one could also investigate $u d$ diquarks. But due
to flavour symmetry (see Eq.~(\ref{eq.two27})) 
 the corresponding matrix elements are proportional 
to those of the $u u$ diquarks  (\ref{dreiquer}) and
(\ref{sechs}). Writing down only the flavour structure one finds for
the expectation values in the proton
\begin{equation}
\langle P | (\ubar \dbar^{\, \rmT}) (d^\rmT u) | P \rangle = 
\frac{1}{4}\langle P | (\ubar \ubar^\rmT) (u^\rmT u) | P \rangle.
\end{equation}

In order to interpret our results we have combined operators from
Tables~\ref{d1} and \ref{d2} such that they correspond to diquarks of
spin zero and spin one.
Specifically, for an operator $\cO_{\mu \nu}$ with two space-time
indices we take the expectation value of $\cO_{44}$ to
represent a spin-zero diquark and the expectation value of
$\sum_{i=1}^3 \cO_{ii}$ to correspond to a spin-one diquark.
The results (once again completely reanalysed) are given in
Table~\ref{dq_spin}. 

\begin{table}
  \caption{Expectation values of operators from the $\mathbf{27}$
    multiplet which correspond to diquarks of spin zero and spin one,
    divided by $m_p^4$ and extrapolated to the chiral limit.
    For the operator $\cO_{\mu\nu}$ the
    spin-zero contribution is $\cO_{44}$, the spin-one contribution is
    $\sum_{i=1}^3 \cO_{ii}$.} \label{dq_spin}
\vspace{0.5cm} \begin{center}
  \begin{tabular}[h]{lcll} \hline
    \multicolumn{1}{c}{operator}  & colour  & 
   \multicolumn{1}{c}{spin 0} & \multicolumn{1}{c}{spin 1} \\ \hline 
    $\frac{1}{10}( \ubar  \gamma_\mu \gamma_5 \gamma_5 C  \ubar^\rmT)
    ( u^\rmT  C^{-1} \gamma_5 \gamma_\nu \gamma_5 u)$&
    $  \overline{\mathbf 3}$ &
    $ (-6.8  \pm  2.4)\cdot 10^{-4}$ & $ (-76  \pm  7)\cdot 10^{-4}$  
    \\
    $\frac{1}{10} (\ubar  \sigma_{\mu \alpha} \gamma_5 C  \ubar^\rmT)
    (u^\rmT C^{-1} \gamma_5   \sigma_{\nu \alpha}  u)$ &
    $  \overline{\mathbf 3}$ &
    $ (-5.7  \pm  6.1)\cdot 10^{-4}$ & $ (48  \pm  18)\cdot 10^{-4}$
    \\
    $\frac{1}{10}(\ubar \gamma_\mu \gamma_5 C  \ubar^\rmT) 
    (\ubar^\rmT C^{-1} \gamma_5  \gamma_\nu u)$ &
    ${\mathbf 6}$ &
    $ (15.0  \pm  3.6)\cdot 10^{-4}$ & $ (30  \pm  7)\cdot 10^{-4}$ \\
   \hline
  \end{tabular}
\vspace{1.0cm} \end{center}
\end{table}

For the operators in the ${\bf \overline{3}}$ of colour the absolute values
for the spin-one diquarks are considerably larger than for the spin-zero
diquarks. For the single operator in the ${\bf 6}$ of colour the difference
is less pronounced. This pattern can tentatively be understood in a
non-relativistic quark picture. When the diquark is in the 
${\bf \overline{3}}$ of SU(3)$_{\mathrm c}$ it is anti-symmetric in the 
colour indices, and therefore the symmetric (in the spin indices) spin-one
state is favoured over the anti-symmetric spin-one state. On the other
hand, when the diquark is in the (symmetric) ${\bf 6}$ of colour one might 
at first sight expect the anti-symmetric spin-zero state to dominate over 
the symmetric spin-one state. Although the spin-zero contribution is indeed
less suppressed than in the ${\bf \overline{3}}$ case it is not really
dominating. This is probably related to the fact that a diquark in the
${\bf 6}$ of SU(3)$_{\mathrm c}$ must be accompanied by (at least) one
gluon if it is to form a colour singlet with the remaining quark.
The coupling to the gluon, mixing ``large'' and ``small'' components
of the quark spinors, would invalidate the above arguments which worked
reasonably well for diquarks in the ${\bf \overline{3}}$ of colour.

Of course, the operators from the ${\mathbf{10}}$ and
${\overline{\mathbf{10}}}$ multiplets can also rewritten in diquark
form. They then appear as linear combinations of operators in
which the diquark is either in a ${\bf \overline{3}}$ of 
SU(3)$_{\mathrm c}$  or in a
${\mathbf{6}}$. The fact that the matrix elements in Table~\ref{t3}
for the colour structures $t_a \otimes t_a$ and $1 \otimes 1$ have
opposite signs translates into a suppression of the ${\mathbf{6}}$
diquarks relative to the  ${\bf \overline{3}}$ diquarks. This is in
accord with our observations made above in the case of the operators
from the  ${\mathbf{27}}$ multiplet.

\section{Summary}

In this paper we have computed the expectation values of a variety of
four-quark operators in the proton by means of quenched Monte Carlo
simulations. Since it is rather difficult to treat the mixing with 
lower-dimensional operators correctly, we have restricted ourselves to
operators whose flavour structure forbids this kind of mixing. The 
additional requirement that the operators should not automatically
vanish in the nucleon led us to consider the flavour group 
SU(3)$_{\mathrm F}$ and to choose operators from the ${\bf 27}$, ${\bf 10}$,
and ${\bf \overline{10}}$ representations. One of the operators from the
${\bf 27}$ is responsible for the twist-four contribution to the lowest
moment of $F_2$ in a somewhat exotic flavour channel. We find a  
rather small value for this contribution. This observation disfavours 
large higher-twist effects in general, although we cannot exclude, of
course, that our result is due to strong cancellations between
different flavour contributions.

Still, and with all due caution, our findings fit into a general trend 
emerging from various pieces of information on higher-twist
contributions to moments of nucleon structure functions: The natural
energy scale for the corresponding correlators lies well below the
nucleon mass leading to small numerical coefficients when the matrix
elements are expressed as multiples of $(m_p)^{t-2}$.

\begin{figure}
  \begin{center}
    \epsfig{file=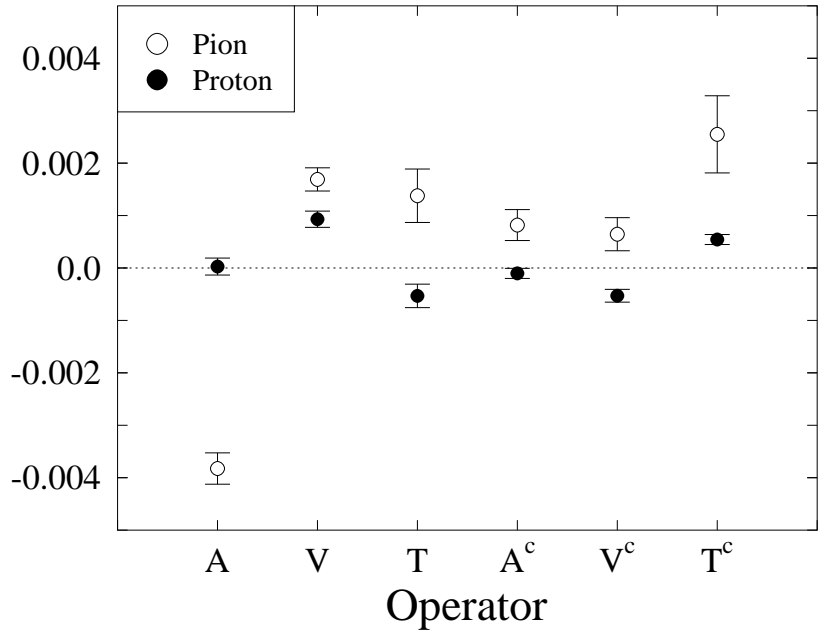,width=14cm} 
    \caption{Renormalised four-quark matrix elements in the pion and in
      the proton (in lattice units). \vspace{1cm}}
    \label{fig.comp}
  \end{center}
\end{figure}

Thus we arrive for the nucleon at a conclusion which is similar to what
we observed in the pion \cite{pion}. For a more detailed comparison we plot in 
Fig.~\ref{fig.comp} the renormalised pion matrix elements \cite{pion} 
with the flavour structure
\begin{equation} \begin{split}
  \langle \pi^+ |(\ubar u)(\ubar u)+(\dbar d)(\dbar d)&-(\ubar u)(\dbar d)
  -(\dbar d)(\ubar u)\\ &-(\ubar d)(\dbar u)-(\dbar u)(\ubar d)
  | \pi^+ \rangle / m_\pi^2 
\end{split}
\end{equation}
together with the corresponding renormalised matrix elements for the proton
$\langle p | 10 \cdot \cO^{I=1}_{\bf 27} | p \rangle /m_p^2$
(in lattice units). We display the results for the spin-two components
setting $\mu = \nu = 4$ (with the trace term subtracted). The normalisation
of the operators is chosen such that the flavour structure 
$(\ubar u)(\ubar u)$ appears with the factor 1 in both cases. 
(Alternatively, it may be remarked that SU(3)$_{\mathrm F}$ makes the 
above pion matrix element equal to the expectation value of 
$10 \cdot \cO^{I=1}_{\bf 27}$ in the meson-octet analogue of the 
proton, the $K^+$.) It is no great surprise that the numbers do not 
show many similarities -- after all, the pion and the proton are very 
different particles.

Four-quark operators in the ${\bf 10}$ and ${\bf \overline{10}}$ 
representations do not lead to much larger matrix elements than the 
${\bf 27}$ operators, although quite a few of those which we studied give
clean signals.
In the ${\bf 27}$ sector, a rewriting of our operators in terms of diquarks
reveals a structure which lends itself to an interpretation with the
help of quark model ideas: Diquarks in the $\overline{\bf 3}$ of 
SU(3)$_{\mathrm c}$ have preferably spin one. Diquarks in the ${\bf 6}$
of SU(3)$_{\mathrm c}$, on the other hand, do not fit so well into a
nonrelativistic picture.

Higher-twist effects will challenge lattice QCD for a few more years. 
Our investigations show that four-quark operators can give reasonable
signals in present quenched Monte Carlo simulations. But the study of 
physically more interesting flavour channels and further twist-four operators
remains an open problem whose solution requires progress in nonperturbative
renormalisation, especially in the treatment of mixing with operators of
lower dimension.

\section*{Acknowledgements}
This work is supported by the DFG (Schwerpunkt ``Elektromagnetische
Sonden'') and by BMBF.
The numerical calculations were performed on the Quadrics 
computers at DESY Zeuthen. We wish to thank the operating staff 
for their support.


\newpage

\begin{thebibliography}{99}
\bibitem{webber}
  S.P. Luttrell, S. Wada and B.R. Webber,
  Nucl.\ Phys.\  B188 (1981) 219; \\
  S.P. Luttrell and S. Wada,
  Nucl.\ Phys.\  B197 (1982) 290; {\it ibid.} B206 (1982) 497~(E).
\bibitem{ellis}
  R.K. Ellis, W. Furma\'nski and R. Petronzio, 
  Nucl.\ Phys.\  B207 (1982) 1; {\it ibid.} B212 (1983) 29.
\bibitem{bag} 
  R.L. Jaffe and M. Soldate, Phys.\ Lett.\ B105 (1981) 467.
\bibitem{wilco}
  R.~L.~Jaffe and M.~Soldate,
  Phys.\ Rev.\ D26 (1982) 49. 
\bibitem{wilco2}
  E.~V.~Shuryak and A.~I.~Vainshtein,
  Nucl.\ Phys.\  B199 (1982) 451.
\bibitem{mira}
  J.L. Miramontes and J. S\'anchez Guill\'en, Z. Phys.\ C41 (1988) 247.
\bibitem{Alekhin:2000ch}
  S.~I.~Alekhin,
  hep-ph/0011002 and private communication.
\bibitem{pion}
  S.~Capitani, M.~G\"ockeler, R.~Horsley, B.~Klaus, V.~Linke, P.E.L.~Rakow,
  A.~Sch\"afer and G.~Schierholz,
  Nucl.\ Phys.\ B570 (2000) 393.
\bibitem{nachtmann}
  O. Nachtmann, Nucl.\ Phys.\  B63 (1973) 237; \\
  S. Wandzura, Nucl.\ Phys.\  B122 (1977) 412.
\bibitem{Gockeler:2000ja}
  M.~G\"ockeler, R.~Horsley, W.~K\"urzinger, H.~Oelrich, D.~Pleiter, 
  P.E.L.~Rakow, A.~Sch\"afer and G.~Schierholz,
  hep-lat/0011091.
\bibitem{Capitani:2000je}
  S.~Capitani, M.~G\"ockeler, R.~Horsley, B.~Klaus, W.~K\"urzinger, 
  D.~Petters, D.~Pleiter, P.E.L.~Rakow, S.~Schaefer, A.~Sch\"afer 
  and G.~Schierholz,
  Nucl.\ Phys.\ B (Proc.\ Suppl.) 94 (2001) 299
\bibitem{Gockeler:1996wg}
  M.~G\"ockeler, R.~Horsley, E.~M.~Ilgenfritz, H.~Perlt, P.~Rakow, 
  G.~Schierholz and A.~Schiller,
  Phys.\ Rev.\ D53 (1996) 2317.
\bibitem{gottlieb}
  S.~Gottlieb, Nucl.\ Phys.\ B139 (1978) 125.
\bibitem{Choi:1993cu}
  S.~Choi, T.~Hatsuda, Y.~Koike and S.~H.~Lee,
  Phys.\ Lett.\ B312 (1993) 351.
\end{thebibliography}
\end{document}